\documentclass[sigconf]{acmart}
\usepackage{array}
\usepackage{caption}
\usepackage{subcaption}
\usepackage{multirow}

\AtBeginDocument{%
  \providecommand\BibTeX{{%
    \normalfont B\kern-0.5em{\scshape i\kern-0.25em b}\kern-0.8em\TeX}}}

\setcopyright{acmcopyright}

\copyrightyear{2019}
\acmYear{2019}
\acmDOI{10.1145/3347448.3357163}

\acmConference[MADiMa '19]{5th International Workshop on Multimedia Assisted Dietary
Management}{October 21, 2019}{Nice, France}
\acmBooktitle{5th International Workshop on Multimedia Assisted Dietary Management
(MADiMa '19), October 21, 2019, Nice, France}
\acmPrice{15.00}
\acmISBN{978-1-4503-6916-9/19/10}

\begin{document}
\fancyhead{}

\title{Self-Attention and Ingredient-Attention Based Model for Recipe Retrieval from Image Queries}


\author{Matthias Fontanellaz}
\email{matthias.fontanellaz@students.unibe.ch}
\affiliation{%
  \institution{ARTORG Center, University of Bern}
  \streetaddress{Murtenstrasse 50}
  \city{Bern}
  \state{Switzerland}
  \postcode{3008}
}

\author{Stergios Christodoulidis}
\email{stergios.christodoulidis@artorg.unibe.ch}
\affiliation{%
  \institution{ARTORG Center, University of Bern}
  \streetaddress{Murtenstrasse 50}
  \city{Bern}
  \state{Switzerland}
  \postcode{3008}
}

\author{Stavroula Mougiakakou}
\email{stavroula.mougiakakou@artorg.unibe.ch}
\affiliation{%
  \institution{ARTORG Center, University of Bern}
  \streetaddress{Murtenstrasse 50}
  \city{Bern}
  \state{Switzerland}
  \postcode{3008}
}

\renewcommand{\shortauthors}{Fontanellaz, et al.}

\begin{abstract}
Direct computer vision based-nutrient content estimation is a demanding task, due to deformation and occlusions of ingredients, as well as high intra-class and low inter-class variability between meal classes. In order to tackle these issues, we propose a system for recipe retrieval from images. The recipe information can subsequently be used to estimate the nutrient content of the meal. In this study, we utilize the multi-modal Recipe1M dataset, which contains over 1 million recipes accompanied by over 13 million images. The proposed model can operate as a first step in an automatic pipeline for the estimation of nutrition content by supporting hints related to ingredient and instruction. Through self-attention, our model can directly process raw recipe text, making the upstream instruction sentence embedding process redundant and thus reducing training time, while providing desirable retrieval results. Furthermore, we propose the use of an ingredient attention mechanism, in order to gain insight into which instructions, parts of instructions or single instruction words are of importance for processing a single ingredient within a certain recipe. Attention-based recipe text encoding contributes to solving the issue of high intra-class/low inter-class variability by focusing on preparation steps specific to the meal. The experimental results demonstrate the potential of such a system for recipe retrieval from images. A comparison with respect to two baseline methods is also presented.
\end{abstract}




\maketitle

\section{Introduction}

Social media and designated online cooking platforms have made it possible for large populations to share food culture (diet, recipes) by providing a vast amount of food-related data. Despite the interest in food culture, global eating behavior still contributes heavily to diet-related diseases and deaths, according to \textit{the Lancet} \cite{FOROUHI20191916}. Nutrition assessment is a demanding, time-consuming and expensive task. Moreover, the conventional approaches for nutrition assessment are cumbersome and prone to errors. A tool that enables users to easily and accurately estimate the nutrition content of a meal, while at the same time minimize the need for tedious work is of great importance for a number of different population groups. Such a tool can be utilized for promoting a healthy lifestyle, as well as to support patients suffering food-related diseases such as diabetes. To this end, a number of computer vision approaches have been developed, in order to extract nutrient information from meal images by using machine learning. Typically, such systems detect the different food items in a picture \cite{DBLP:journals/corr/abs-1711-05128}, \cite{7169816}, \cite{8181494}, estimate their volumes \cite{DBLP:journals/corr/abs-1806-10343}, \cite{7792736}, \cite{6607548} and calculate the nutrient content using a food composition database \cite{FoodData}. In some cases however, inferring the nutrient content of a meal from an image can be really challenging - due to unseen ingredients (e.g. sugar, oil) or the structure of the meal (mixed food, soups, etc.). 

Humans often use information from diverse sensory modalities (visual, auditory, haptic) to infer logical conclusions. This kind of multi-sensory integration helps us process complex tasks \cite{MultisensoryIntegration}. In this study, we investigate the use of recipe information, in order to better estimate nutrient content of complex meal compositions. With the aim to develop a pipeline for holistic dietary assessment, we present and evaluate a method based on machine learning to retrieve recipe information from images, as a first step towards more accurate nutrient estimation. Such recipe information can then be utilized together with the volume of the food item to enhance an automatic system to estimate the nutrient content of complex meals, such as lasagna, crock pot or stew. 

\begin{figure*}
    \centering
    \includegraphics[width=0.95\textwidth]{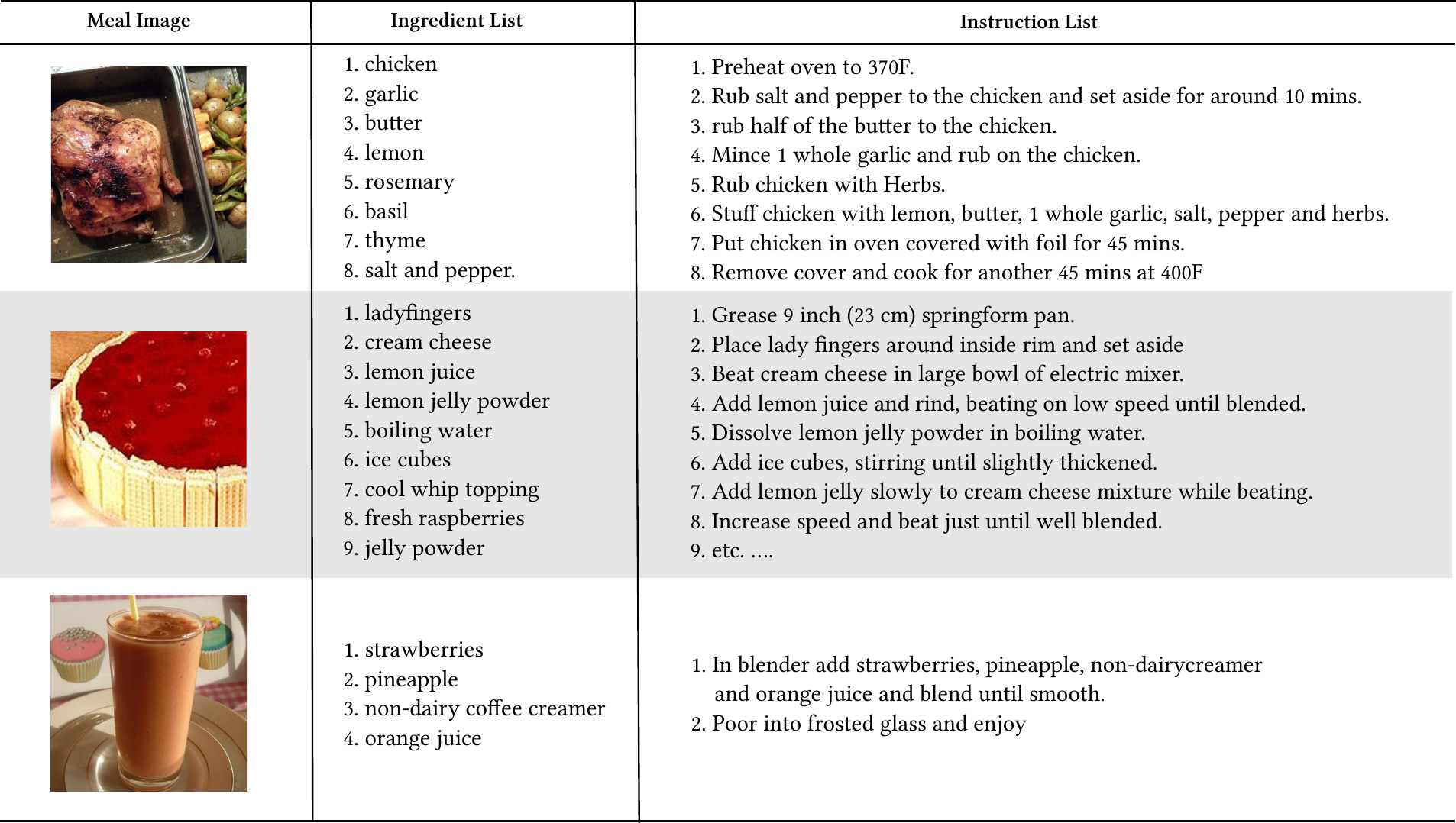}
    \caption{Recipe samples from the Recipe1M Dataset.}
    \label{fig:recipes}
\end{figure*}

The performance of approaches based on machine learning relies heavily on the quantity and quality of the available data. To this end, a number of efforts have been made to compile informative datasets to be used for machine learning approaches. Most of the early released food databases were assembled only by image data for a special kind of meal. In particular, the first publicly available database was the Pittsburgh Fast-Food Image Dataset (PFID) \cite{pfid}, which contains only fast food images taken under laboratory conditions. After the recent breakthrough in deep learning models, a number of larger databases were introduced. Bossard et al. \cite{Food-101} introduced the Food-101 dataset, which is composed of 101 food categories represented by 101'000 food images. This was followed by several image-based databases, such as the UEC-100 \cite{matsuda12} and its augmented version, the UEC-256 \cite{kawano14c} dataset, with 9060 food images referring to 100 Japanese food types and 31651 food images referring to 256 Japanese food types, respectively. Xu et al. \cite{GeolocalizedModelingforDishRecognition} developed a specialized dataset by including geolocation and external information about restaurants to simplify the food recognition task. Wang et al. \cite{Wang} introduced the UPMC Food-101 multi-modal dataset, that shares the same 101 food categories with the popular Food-101 dataset, but contains textual information in addition. A number of studies have been carried out utilizing the aforementioned databases, mainly for the task of food recognition. Salvador et al. \cite{Salvador} published Recipe1M, the largest publicly available multi-modal dataset, that consists of 1 million recipes together with the accompanying images.

The emergence of multi-modal databases has led to novel approaches for meal image analysis. The fusion of visual features learned from images by deep Convolution Neural Networks (CNN) and textual features lead to outstanding results in food recognition applications. An early approach for recipe retrieval was based on jointly learning to predict food category and its ingredients using deep CNN \cite{IngredientRecognition}. In a following step, the predicted ingredients are matched against a large corpus of recipes. More recent approach is proposed by \cite{Salvador} and is based on jointly learning recipe-text and image representations in a shared latent space. Recurrent Neural Networks (RNN) and CNN are mainly used to map text and image into the shared space. To align the text and image embedding vectors between matching recipe-image pairs, cosine similarity loss with  margin was applied. Carvalho et al. \cite{Carvalho} proposed a similar multi-modal embedding method for aligning text and image representations in a shared latent space. In contrast to Salvador et al. \cite{Salvador}, they formulated a joint objective function which incorporates the loss for the cross-modal retrieval task and a classification loss, instead of using the latent space for a multitask learning setup. To address the challenge of encoding long sequences (like recipe instructions), \cite{Salvador} chose to represent single instructions as sentence embedding using the skip-thought technique \cite{DBLP:journals/corr/KirosZSZTUF15}. These encoded instruction sentences are referred to as skip-instructions and their embedding is not fine tuned when learning the image-text joint embedding.

In this study, we present a method for the joint learning of meal image and recipe embedding, using a multi-path structure that incorporates natural language processing paths, as well as image analysis paths. The main contribution of the proposed method is threefold: i) the direct encoding of the instructions, ingredients and images during training, making the need of skip instruction embedding redundant; ii) the utilization of multiple attention mechanisms (i.e. self-attention and ingredient-attention), and iii) a lightweight architecture.

\section{Materials and Methods}

\subsection{Database}

The proposed method is trained and evaluated on Recipe1M \cite{Salvador}, the largest publicly available multi-modal food database. Recipe1M provides over 1 million recipes (ingredients and instructions), accompanied by one or more images per recipe, leading to 13 million images. The large corpus is supplemented with semantic information (1048 meal classes) for injecting an additional source of information in potential models. In the table in Figure \ref{fig:recipes},  the structure of recipes belonging to different semantic classes is displayed. Using a slightly adjusted pre-processing than that in \cite{Salvador} (elimination of noisy instruction sentences), the training set, validation set and test set contain 254,238 and 54,565 and 54,885 matching pairs, respectively. In \cite{Salvador}, the authors chose the overall amount of instructions per recipe as one criterion for a valid matching pair. But we simply removed instruction sentences that contain only punctuation and gained some extra data for training and validation.

\subsection{Model Architecture}

\begin{figure}
    \centering
    \includegraphics[width=0.45\textwidth]{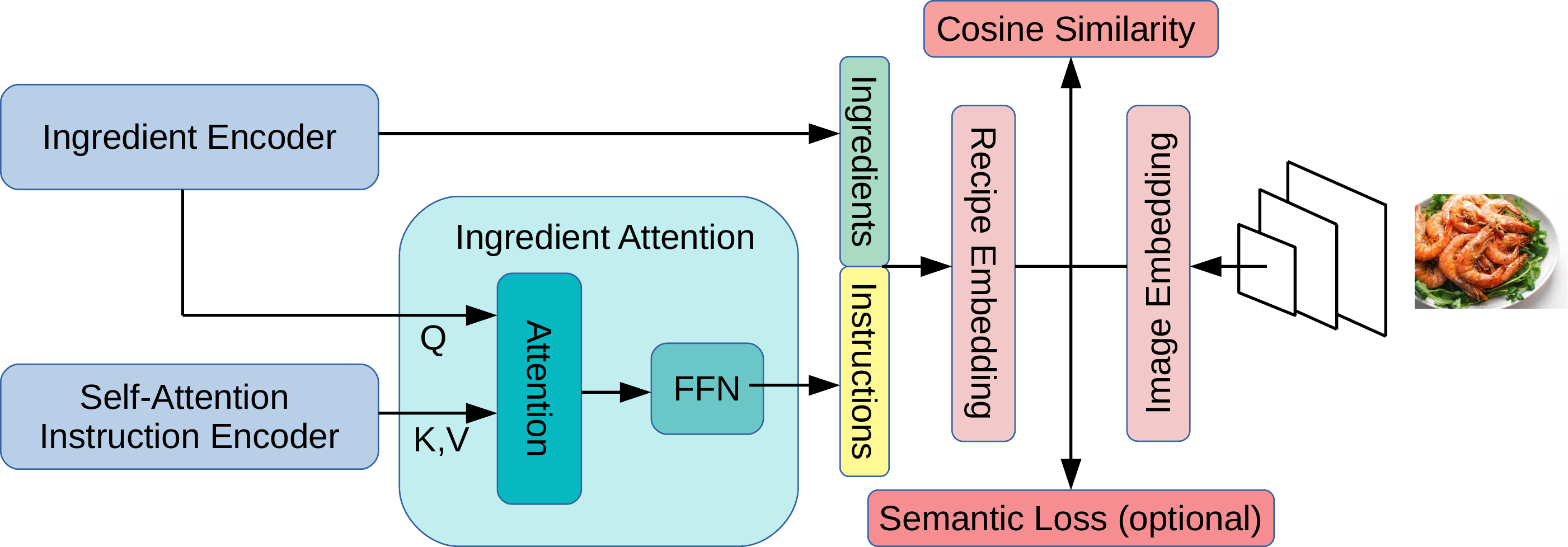}
    \caption{Text-image embedding model with optional semantic classifier for semantic regularization according to \cite{Recipe1M} and with Ingredient Attention based instruction encoding}
    \label{fig:EmbeddingModel}
\end{figure}

The proposed model architecture is based on a multi-path approach for each of the involved input data types namely, instructions, ingredients and images, similarly to \cite{Recipe1M}. In Figure \ref{fig:EmbeddingModel}, the overall structure is presented. For the instruction encoder, we utilized a self-attention mechanism \cite{DBLP:journals/corr/VaswaniSPUJGKP17}, which learns which words of the instructions are relevant with a certain ingredient. In order to encode the ingredients, a bidirectional RNN is used, since ingredients are an unordered list of words. All RNNs in the ingredients path were implemented with Long Short-Term Memory (LSTM) cells \cite{Hochreiter:1997:LSM:1246443.1246450}. We fixed the ingredient representation to have a length of 600, independent of the amount of ingredients. Lastly, the outputs of the self-attention-instruction encoder with ingredient attention and the output of the bidirectional LSTM ingredient-encoder are concatenated and mapped to the joint embedding space. The image analysis path is composed of a ResNet-50 model \cite{DBLP:journals/corr/HeZRS15}, pretrained on the ImageNet Dataset \cite{imagenet_cvpr09}, with a custom top layer for mapping the image features to the joint embedding space. All word embeddings are pretrained with the word2vec algorithm \cite{Mikolov:2013:DRW:2999792.2999959} and fine tuned during the joint embedding learning phase. We chose 512-dimensional word embedding for our model with self-attention, whereas \cite{Recipe1M} and \cite{Carvalho} chose a vector length of 300. In the following sections, more details about the aforementioned paths are presented.  

\subsection{Attention Mechanisms}

The instruction encoder follows a transformer based encoder, as suggested by \cite{DBLP:journals/corr/VaswaniSPUJGKP17}. Since we do not focus on syntactic rules, but mostly on weak sentence semantics or single words, we built a more shallow encoder containing only 2 stacked layers, where each of this layers contains two sub-layers. The first is the multi-head attention layer, and the second is a position-wise densely connected feed-forward network (FFN). Due to recipes composed of over 600 words as instructions, we decided to trim words per instruction sentence to restrict the overall words per recipe to 300. In order to avoid removing complete instructions at the end of the instruction table, we removed a fraction of words from each instruction, based on this instruction's length and the overall recipe-instruction length.
This strategy reinforces the neglect of syntactic structures in the instruction encoding process. With such a model, we can directly perform the instruction encoding during the learning process for the joint embedding, thus saving training time and reducing disk space consumption. The transformer-like encoder does not make use of any recurrent units, thus providing the opportunity for a more lightweight architecture. By using self-attention \cite{DBLP:journals/corr/VaswaniSPUJGKP17}, the model learns to focus on instructions relevant to recipe-retrieval-relevant, parts of instructions or single instruction-words. Furthermore we gain insight into which instructions are important to distinguish recipes with similar ingredients but different preparation styles.

The instruction encoder transforms the sequence of plain word representations with added positional information to a sequence of similarity-based weighted sum of all word representations. The outputted sequence of the encoder exhibits the same amount of positions as the input to the instruction encoder (in our experiments 300). Each of this positions is represented by a 512-dimensional vector. To obtain a meaningful representation without a vast number of parameters, we reduced the number of word representations before the concatenation with the ingredient representation. For this reduction step, we implemented a recipe-embedding specific attention layer where the ingredient representation is used to construct $n$ queries, where $n$ is the amount of new instruction representation vectors. Each of these new representations is a composition of all previous word representations weighted by the ingredient attention score. Following, the ingredient attention process is formulated mathematically and is visually portrayed in Figure \ref{fig:EmbeddingModel}.
\begin{equation}
    IA(K(inst),V(inst),Q(ing))=Sofmax\bigg(\frac{K(inst)*Q(ing)^T}{\sqrt{d_k}}\bigg)
\end{equation}
where $K(inst)$ and $V(inst)$ are linear mappings of the encoded instruction words, and $Q(ing)$ is a linear mapping of the ingredient representation and $d_k$ is the dimensionality of linearly projected position vectors. 
\begin{equation*}
    K(inst)=inst*W_k,\:with\:inst\in\!R^{bxpxw}\:and\:W_k\in\!R^{wxh}
\end{equation*}

\begin{equation*}
    V(inst)=inst*W_v,\:with\:inst\in\!R^{bxpxw}\:and\:W_v\in\!R^{wxw}
\end{equation*}

\begin{equation*}
    Q(ing)=ing*W_q,\:with\:inst\in\!R^{bxq}\:and\:W_q\in\!R^{nxqxh}
\end{equation*}
where $b$ is the batch-size, $p$ is the amount of word embeddings, $w$ is the dimensionality of the wort embedding, $h$ is the dimensionality of the space to where we project the word embeddings and queries, $q$ is the dimensionality of the ingredient representation and $n$ is the amount of Ingredient Attention-based instruction representations. Ingredient Attention can be performed step-wise, similarly to the well known dimensionality reduction in convolution neural networks.

\subsection{Loss function}
To align text and image embeddings of matching recipe-image pairs alongside each other, we maximize the cosine distance between positive pairs and minimize it between negative pairs.

We have trained our model using cosine similarity loss with margin as in \cite{Recipe1M} and with the triplet loss proposed by \cite{Carvalho}. Both objective functions and the semantic regularization by \cite{Recipe1M} aim at maximizing intra-class correlation and minimizing inter-class correlation. 

Let us define the text query embedding as $\phi^q$ and the embedding of the image query as $\phi^d$, then the cosine embedding loss can be defined as follows:

\begin{equation*}
    L_{cos}(\phi^q,\phi^d,y)=\left\{\begin{array}{ll} 1-\cos(\phi^q,\phi^d), & if \; y=1 \\
                    \max(0, \cos(\phi^q,\phi^d)-\alpha, & if \; y=-1\end{array}\right.
\end{equation*}

where $cos(x,y)$ is the normalized cosine similarity and $\alpha$ is a margin ($-1\leqslant \alpha \leqslant 1)$, that determines how similar negative pairs are allowed to be. Positive margins allow negative pairs to share at maximum $\alpha$ similarity, where a maximum margin of zero or negative margins allow no correlation between non matching embedding vectors or force the model to learn antiparallel representations, respectively. $\phi^d$ is the corresponding image counterpart to $\phi^q$ if $y=1$ or a randomly chosen sample $\phi^d \in S \wedge \phi^d \ne \phi^{d(q)}$ if $y=-1$, where $\phi^{d(q)}$ is the true match for $\phi^q$ and $S$ is the dataset we sample from it. Furthermore, we complement the cosine similarity with cross-entropy classification loss ($L_{reg}$), leading to the applied objective function.

\begin{equation}
    L(\phi^q,\phi^d,c_r,c_v,y)=L_{cos}(\phi^q,\phi^d,y)+\lambda*L_{reg}(\phi^q,\phi^d,c_r,c_v)
\end{equation}

with $c_r$ and $c_v$ as semantic recipe-class and semantic image-class, respectively, while $c_r=c_v$ if the food image and recipe text are a positive pair.

For the triplet loss, we define $\phi^q$ as query embedding, $\phi^{d+}$ as matching image counterpart and $\phi^{d-}$ as another random sample taken from $S$. Further $\phi^{d_{sem}+} \in S \wedge \phi^{d_{sem}+} \ne \phi^{d(q)}$ is a sample from $S$ sharing the same semantic class as $\phi^q$ and $\phi^{d_{sem}-}$ is a sample from any other class. The triplet loss is formulated as follows:

\begin{equation*}
    L_{sample}(\phi^q, \phi^{d+},\phi^{d-})=\big[\alpha - \cos(\phi^q,\phi^{d+}) + \cos(\phi^q,\phi^{d-})\big]
\end{equation*}
\begin{multline*}
    L_{sem}(\phi^q,\phi^{d_{sem}+},\phi^{d_{sem}-})=\\ \big[\alpha - \cos(\phi^q,\phi^{d_{sem}+}) + \cos(\phi^q,\phi^{d_{sem}-})\big]
\end{multline*}

\begin{multline}
    L_{triplet}(\phi^q, \phi^{d+},\phi^{d-},\phi^{d_{sem}+},\phi^{d_{sem}-})=\\ \beta*{ L_{sample}(\phi^q, \phi^{d+},\phi^{d-})}^2 + \\ (1-\beta)*\big( L_{sample}(\phi^q, \phi^{d+},\phi^{d-})\big) + \\
    \gamma*\Big(\beta*{L_{sem}(\phi^q,\phi^{d_{sem}+},\phi^{d_{sem}-})}^2 + \\ (1-\beta)*\big(L_{sem}(\phi^q,\phi^{d_{sem}+},\phi^{d_{sem}-})\big)\Big)
\end{multline}

where $\beta\in[0,1]$ weights between quadratic and linear loss, $\alpha\in [0,2]$ is the margin and $\gamma\in [0,1]$ weights between semantic- and sample-loss. The triplet loss encourages the embedding vectors of a matching pair to be larger by a margin above its non-matching counterpart. Further, the semantic loss encourages the model to form clusters of dishes, sharing the same class. We chose  $\beta$ to be $0.1$, $\alpha$ to be $0.3$ and $\gamma$ to be $0.3$.

\subsection{Training configuration}

We used Adam \cite{Adam} optimizer with an initial learning rate of $10^{-4}$. At the beginning of the training session, we freeze the pretrained ResNet-50 weights and optimize only the text-processing branch until we do no longer make progress. Then, we alternate train image and text branch until we switched modality for 10 times. Lastly, we fine-tune the overall model by releasing all trainable parameters in the model. Our optimization strategy differs from \cite{Recipe1M} in that we use an aggressive learning rate decay, namely exponential decay, so that the learning rate is halved all 20 epochs. Since the timing of freezing layers proved not to be of importance unless the recipe path is trained first, we used the same strategy under the cosine distance objective \cite{Recipe1M} and for the triplet loss \cite{Carvalho}.

\section{Experimental Setup and Results}

Recipe1M is already distributed in three parts, the training, validation and testing sets. We did not make any changes to these partitions. Except with our more sensitive preprocessing algorithm, we accept more recipes from the raw corpus. \cite{Recipe1M} used 238,399 samples for their effective training set and for the validation and testing set 51,119 and 51,303 samples, respectively. By filtering out noisy instructions sentences (e.g. instructions containing only punctuation) we increased the effective dataset size to 254,238 samples for the training set and 54,565 and 54,885 for the validation and testing sets, respectively.

\begin{table}
\caption{Comparison between our method, our Joint Neural Embedding (JNE)\cite{Recipe1M} and AdaMine \cite{Carvalho} re-implementation. For all models we were using selected matching pairs generated by reducing noisy instruction sentences as described above. Recall rates are averaged over the evaluation batches.}
\begin{tabular}{|c|c|c|c|c|c|}
 \hline
 \multicolumn{6}{|c|}{Image to Recipe} \\
 \hline
 & & MedR & R@1 & R@5 & R@10\\ [0.3ex] 
 \hline
 \multirow{4}{*}{\rotatebox[origin=c]{90}{1k samples}} & Random \cite{Recipe1M} & $500.0$ & $0.001$ & $0.005$ & $0.01$\\ 
 & JNE \cite{Recipe1M}& $5.0\pm0.1$ & $25.9$ & $52.6$ & $64.1$\\
 & AdaMine \cite{Carvalho}& $3.0\pm0.1$ & $33.1$ & $64.3$ & $75.2$\\
 & IA & $2.9\pm0.3$ & $34.6$ & $66.0$ & $76.6$\\
 \hline
\end{tabular}
\label{table:results}
\end{table}

Similarly to \cite{Recipe1M} and \cite{Carvalho}, we evaluated our model on 10 subsets of 1000 samples each. One sample of these subsets is composed of text embedding and image embedding in the shared latent space. Since our interest lies in the recipe retrieval task, we optimized and evaluated our model by using each image embedding in the subsets as query against all text embeddings. By ranking the query and the candidate embeddings according to their cosine distance, we estimate the median rank. The model's performance is best, if the matching text embedding is found at the first rank. Further, we estimate the recall percentage at the top K percent over all queries. The recall percentage describes the quantity of queries ranked amid the top K closest results. In Table \ref{table:results} the results are presented, in comparison to baseline methods.

\begin{figure}[ht]
\centering
   \begin{subfigure}{0.45\textwidth}
   \includegraphics[width=\textwidth]{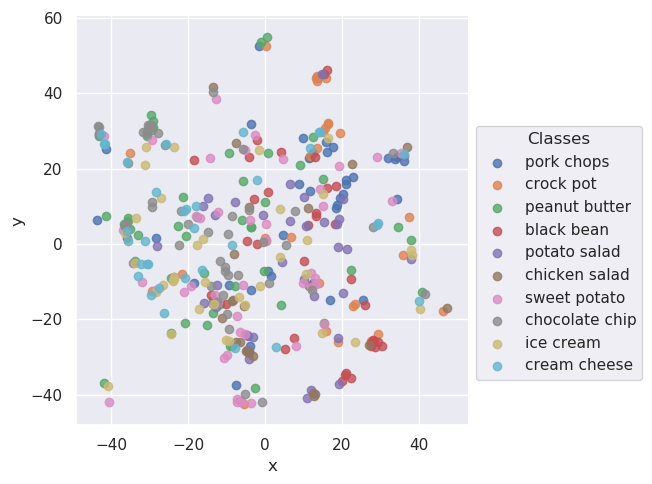}
   \caption{}
   \label{fig:tSNE_cos} 
\end{subfigure}
\begin{subfigure}{0.45\textwidth}
   \includegraphics[width=\textwidth]{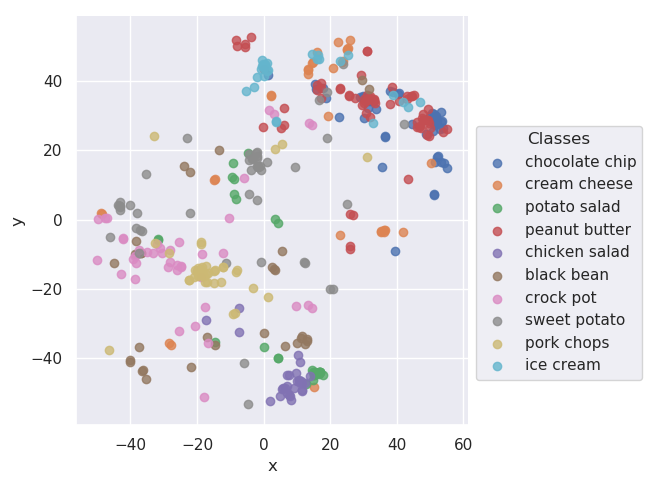}
   \caption{}
   \label{fig:tSNE_triplet}
\end{subfigure}
\caption{(a) Visualization of the joint embedding space under the cosine distance with semantic regularization objective. (b) organization of the joint embedding space under the triplet}
\label{fig:tSNE}
\end{figure}

Both \cite{Recipe1M} and \cite{Carvalho} use time-consuming instruction text preprocessing over the skip-thought technique \cite{DBLP:journals/corr/KirosZSZTUF15}. This process doubles the overall training time from three days to six days using two Nvidia Titan X GPU's. By using online-instruction encoding with the self-attention encoder, we were able train the model for its main task in under 30 hours. Furthermore, the proposed approach offers more flexibility for dataset alterations. 

\begin{figure}[ht]
  \centering
  \includegraphics[width=0.49\textwidth]{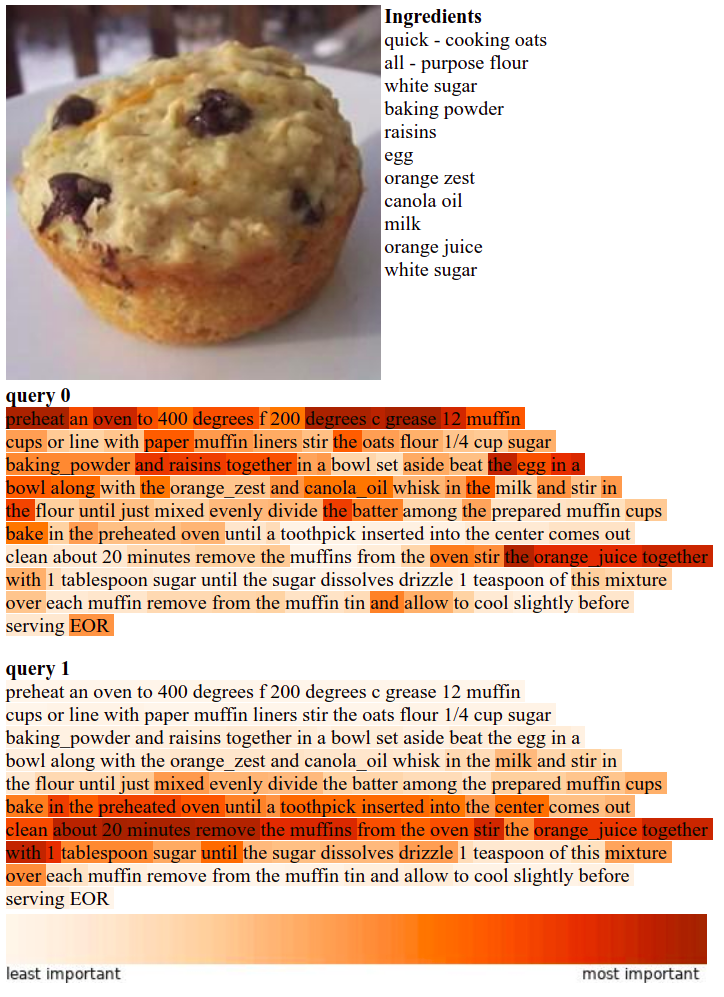}
  \caption{Ingredient-Attention based focus on instruction sentences. We use two different mapping matrices for the two ingredient based queries.}
  \label{fig:HeatMap}
\end{figure}

\begin{figure*}
    \centering
    \includegraphics[width=0.95\textwidth]{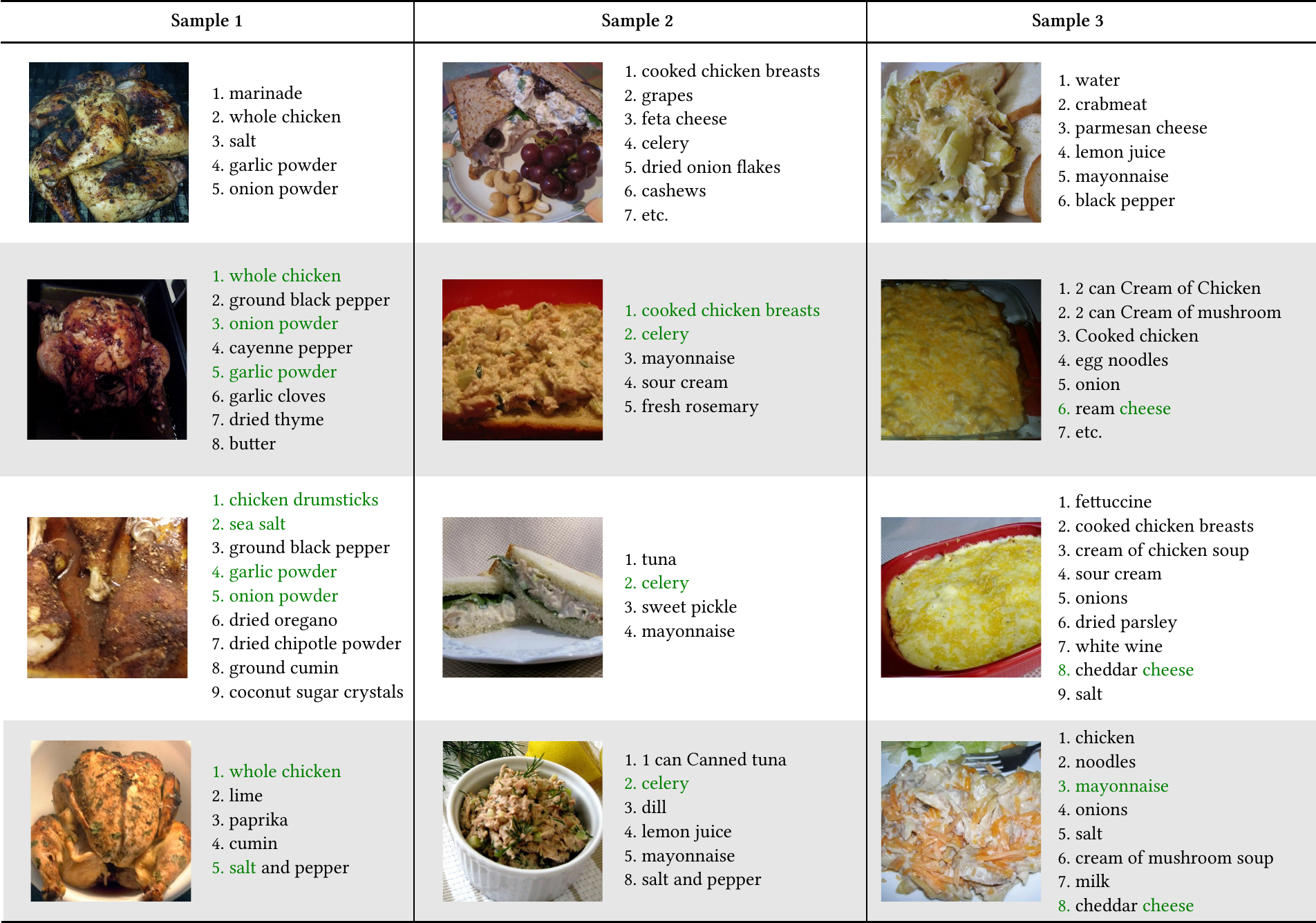}
    \caption{The retrieval performance of our model depends heavily on the meal type. We marked matching retrieved ingredients or those of the same family in green. The Ingredient Attention model performed well on Sample 1, and acceptably on Sample 2. On Sample 3, the model missed the main ingredient in all top three retrievals.}
    \label{fig:retrievals}
\end{figure*}

Qualitative results such as recipe retrieval, quality of the cluster formation in the joint embedding space and heat maps of instruction words are more important than the previously mentioned benchmarking scores. Depending on meal type, all baseline implementations as well as our Ingredient Attention based model exhibit a broad range of retrieval accuracy. In Figure \ref{fig:retrievals} we present a few typical results on the intended recipe retrieval task.

AdaMine \cite{Carvalho} creates more distinct class clusters than in \cite{Recipe1M}. In Figure \ref{fig:tSNE}, we demonstrate the difference in cluster formation using the aforementioned Methods for our Ingredient Attention. We visualize the top ten most common recipe classes in Recipe1M using t-SNE \cite{Maaten08visualizingdata}. Since chocolate chip, peanut butter, cream cheese and/or ice cream are used as ingredients in desserts, due to semantic regularization inside the triplet loss, clusters of sweet meals are close together (Figure \ref{fig:tSNE_triplet} top right corner).

We use heat maps on instruction words as tool to visualize words relevant to ingredient-lists in plain instruction text. In Figure \ref{fig:HeatMap}, we demonstrate how easily we can achieve insight into the models decision making.

\section{Conclusions}

In this paper, we have introduced self-attention for instruction encoding in the context of the recipe retrieval task and ingredient attention for disclosing ingredient dependent meal preparation steps. Our main contribution is the aforementioned ingredient attention, empowering our model to solve the recipe retrieval without any upstream skip instruction embedding, as well as the light-weight architecture provided by the transformer-like instruction encoder. On the recipe retrieval task, our method performs similarly to our baseline implementation of \cite{Carvalho}. Regarding training time on the other hand, we increased the efficiency significantly for cross-modal based retrieval methods. There is no need for a maximum number of instructions for a recipe to be considered as valid for training or testing; only for total words, making more samples of the large Recipe1M corpus usable for training. Through ingredient attention, we are able to unveil internal focus in the text processing path by observing attention weights. Incorporation of new samples in the train set can be done by retraining just one model. Overall, an accurate and  flexible method for recipe retrieval from meal images could provide downstream models (e.g. automatic nutrient content estimation) with decisive information and significantly improve their results.

\bibliographystyle{ACM-Reference-Format}
\bibliography{sample-base}


\begin{thebibliography}{27}


\ifx \showCODEN    \undefined \def \showCODEN     #1{\unskip}     \fi
\ifx \showDOI      \undefined \def \showDOI       #1{#1}\fi
\ifx \showISBNx    \undefined \def \showISBNx     #1{\unskip}     \fi
\ifx \showISBNxiii \undefined \def \showISBNxiii  #1{\unskip}     \fi
\ifx \showISSN     \undefined \def \showISSN      #1{\unskip}     \fi
\ifx \showLCCN     \undefined \def \showLCCN      #1{\unskip}     \fi
\ifx \shownote     \undefined \def \shownote      #1{#1}          \fi
\ifx \showarticletitle \undefined \def \showarticletitle #1{#1}   \fi
\ifx \showURL      \undefined \def \showURL       {\relax}        \fi
\providecommand\bibfield[2]{#2}
\providecommand\bibinfo[2]{#2}
\providecommand\natexlab[1]{#1}
\providecommand\showeprint[2][]{arXiv:#2}

\bibitem[\protect\citeauthoryear{Aguilar, Remeseiro, Bola{\~{n}}os, and
  Radeva}{Aguilar et~al\mbox{.}}{2017}]%
        {DBLP:journals/corr/abs-1711-05128}
\bibfield{author}{\bibinfo{person}{Eduardo Aguilar}, \bibinfo{person}{Beatriz
  Remeseiro}, \bibinfo{person}{Marc Bola{\~{n}}os}, {and}
  \bibinfo{person}{Petia Radeva}.} \bibinfo{year}{2017}\natexlab{}.
\newblock \showarticletitle{Grab, Pay and Eat: Semantic Food Detection for
  Smart Restaurants}.
\newblock \bibinfo{journal}{\emph{CoRR}}  \bibinfo{volume}{abs/1711.05128}
  (\bibinfo{year}{2017}).
\newblock
\showeprint[arxiv]{1711.05128}
\urldef\tempurl%
\url{http://arxiv.org/abs/1711.05128}
\showURL{%
\tempurl}


\bibitem[\protect\citeauthoryear{Bossard, Guillaumin, and Van~Gool}{Bossard
  et~al\mbox{.}}{2014}]%
        {Food-101}
\bibfield{author}{\bibinfo{person}{Lukas Bossard}, \bibinfo{person}{Matthieu
  Guillaumin}, {and} \bibinfo{person}{Luc Van~Gool}.}
  \bibinfo{year}{2014}\natexlab{}.
\newblock \showarticletitle{Food-101 -- Mining Discriminative Components with
  Random Forests}. In \bibinfo{booktitle}{\emph{Computer Vision -- ECCV 2014}},
  \bibfield{editor}{\bibinfo{person}{David Fleet}, \bibinfo{person}{Tomas
  Pajdla}, \bibinfo{person}{Bernt Schiele}, {and} \bibinfo{person}{Tinne
  Tuytelaars}} (Eds.). \bibinfo{publisher}{Springer International Publishing},
  \bibinfo{address}{Cham}, \bibinfo{pages}{446--461}.
\newblock
\showISBNx{978-3-319-10599-4}


\bibitem[\protect\citeauthoryear{Carvalho, Cad{\`{e}}ne, Picard, Soulier,
  Thome, and Cord}{Carvalho et~al\mbox{.}}{2018}]%
        {Carvalho}
\bibfield{author}{\bibinfo{person}{Micael Carvalho},
  \bibinfo{person}{R{\'{e}}mi Cad{\`{e}}ne}, \bibinfo{person}{David Picard},
  \bibinfo{person}{Laure Soulier}, \bibinfo{person}{Nicolas Thome}, {and}
  \bibinfo{person}{Matthieu Cord}.} \bibinfo{year}{2018}\natexlab{}.
\newblock \showarticletitle{Cross-Modal Retrieval in the Cooking Context:
  Learning Semantic Text-Image Embeddings}.
\newblock \bibinfo{journal}{\emph{CoRR}}  \bibinfo{volume}{abs/1804.11146}
  (\bibinfo{year}{2018}).
\newblock
\showeprint[arxiv]{1804.11146}
\urldef\tempurl%
\url{http://arxiv.org/abs/1804.11146}
\showURL{%
\tempurl}


\bibitem[\protect\citeauthoryear{Chen and Ngo}{Chen and Ngo}{2016}]%
        {IngredientRecognition}
\bibfield{author}{\bibinfo{person}{Jingjing Chen} {and}
  \bibinfo{person}{Chong-Wah Ngo}.} \bibinfo{year}{2016}\natexlab{}.
\newblock \showarticletitle{Deep-based Ingredient Recognition for Cooking
  Recipe Retrieval}. \bibinfo{pages}{32--41}.
\newblock
\urldef\tempurl%
\url{https://doi.org/10.1145/2964284.2964315}
\showDOI{\tempurl}


\bibitem[\protect\citeauthoryear{Chen, Dhingra, Wu, Yang, Sukthankar, and
  Yang}{Chen et~al\mbox{.}}{[n.d.]}]%
        {pfid}
\bibfield{author}{\bibinfo{person}{Mei Chen}, \bibinfo{person}{Kapil Dhingra},
  \bibinfo{person}{Wen Wu}, \bibinfo{person}{Lei Yang}, \bibinfo{person}{Rahul
  Sukthankar}, {and} \bibinfo{person}{Jie Yang}.}
  \bibinfo{year}{[n.d.]}\natexlab{}.
\newblock \bibinfo{title}{PFID: PITTSBURGH FAST-FOOD IMAGE DATASET}.
\newblock
\newblock


\bibitem[\protect\citeauthoryear{{Dehais}, {Anthimopoulos}, {Shevchik}, and
  {Mougiakakou}}{{Dehais} et~al\mbox{.}}{2017}]%
        {7792736}
\bibfield{author}{\bibinfo{person}{J. {Dehais}}, \bibinfo{person}{M.
  {Anthimopoulos}}, \bibinfo{person}{S. {Shevchik}}, {and} \bibinfo{person}{S.
  {Mougiakakou}}.} \bibinfo{year}{2017}\natexlab{}.
\newblock \showarticletitle{Two-View 3D Reconstruction for Food Volume
  Estimation}.
\newblock \bibinfo{journal}{\emph{IEEE Transactions on Multimedia}}
  \bibinfo{volume}{19}, \bibinfo{number}{5} (\bibinfo{date}{May}
  \bibinfo{year}{2017}), \bibinfo{pages}{1090--1099}.
\newblock
\showISSN{1520-9210}
\urldef\tempurl%
\url{https://doi.org/10.1109/TMM.2016.2642792}
\showDOI{\tempurl}


\bibitem[\protect\citeauthoryear{Deng, Dong, Socher, Li, Li, and Fei-Fei}{Deng
  et~al\mbox{.}}{2009}]%
        {imagenet_cvpr09}
\bibfield{author}{\bibinfo{person}{J. Deng}, \bibinfo{person}{W. Dong},
  \bibinfo{person}{R. Socher}, \bibinfo{person}{L.-J. Li}, \bibinfo{person}{K.
  Li}, {and} \bibinfo{person}{L. Fei-Fei}.} \bibinfo{year}{2009}\natexlab{}.
\newblock \showarticletitle{{ImageNet: A Large-Scale Hierarchical Image
  Database}}. In \bibinfo{booktitle}{\emph{CVPR09}}.
\newblock


\bibitem[\protect\citeauthoryear{Forouhi and Unwin}{Forouhi and Unwin}{2019}]%
        {FOROUHI20191916}
\bibfield{author}{\bibinfo{person}{Nita~G Forouhi} {and} \bibinfo{person}{Nigel
  Unwin}.} \bibinfo{year}{2019}\natexlab{}.
\newblock \showarticletitle{Global diet and health: old questions, fresh
  evidence, and new horizons}.
\newblock \bibinfo{journal}{\emph{The Lancet}} \bibinfo{volume}{393},
  \bibinfo{number}{10184} (\bibinfo{year}{2019}), \bibinfo{pages}{1916 --
  1918}.
\newblock
\showISSN{0140-6736}
\urldef\tempurl%
\url{https://doi.org/10.1016/S0140-6736(19)30500-8}
\showDOI{\tempurl}


\bibitem[\protect\citeauthoryear{He, Zhang, Ren, and Sun}{He
  et~al\mbox{.}}{2015}]%
        {DBLP:journals/corr/HeZRS15}
\bibfield{author}{\bibinfo{person}{Kaiming He}, \bibinfo{person}{Xiangyu
  Zhang}, \bibinfo{person}{Shaoqing Ren}, {and} \bibinfo{person}{Jian Sun}.}
  \bibinfo{year}{2015}\natexlab{}.
\newblock \showarticletitle{Deep Residual Learning for Image Recognition}.
\newblock \bibinfo{journal}{\emph{CoRR}}  \bibinfo{volume}{abs/1512.03385}
  (\bibinfo{year}{2015}).
\newblock
\showeprint[arxiv]{1512.03385}
\urldef\tempurl%
\url{http://arxiv.org/abs/1512.03385}
\showURL{%
\tempurl}


\bibitem[\protect\citeauthoryear{{He}, {Xu}, {Khanna}, {Boushey}, and
  {Delp}}{{He} et~al\mbox{.}}{2013}]%
        {6607548}
\bibfield{author}{\bibinfo{person}{Y. {He}}, \bibinfo{person}{C. {Xu}},
  \bibinfo{person}{N. {Khanna}}, \bibinfo{person}{C.~J. {Boushey}}, {and}
  \bibinfo{person}{E.~J. {Delp}}.} \bibinfo{year}{2013}\natexlab{}.
\newblock \showarticletitle{Food image analysis: Segmentation, identification
  and weight estimation}. In \bibinfo{booktitle}{\emph{2013 IEEE International
  Conference on Multimedia and Expo (ICME)}}. \bibinfo{pages}{1--6}.
\newblock
\showISSN{1945-7871}
\urldef\tempurl%
\url{https://doi.org/10.1109/ICME.2013.6607548}
\showDOI{\tempurl}


\bibitem[\protect\citeauthoryear{Hochreiter and Schmidhuber}{Hochreiter and
  Schmidhuber}{1997}]%
        {Hochreiter:1997:LSM:1246443.1246450}
\bibfield{author}{\bibinfo{person}{Sepp Hochreiter} {and}
  \bibinfo{person}{J\"{u}rgen Schmidhuber}.} \bibinfo{year}{1997}\natexlab{}.
\newblock \showarticletitle{Long Short-Term Memory}.
\newblock \bibinfo{journal}{\emph{Neural Comput.}} \bibinfo{volume}{9},
  \bibinfo{number}{8} (\bibinfo{date}{Nov.} \bibinfo{year}{1997}),
  \bibinfo{pages}{1735--1780}.
\newblock
\showISSN{0899-7667}
\urldef\tempurl%
\url{https://doi.org/10.1162/neco.1997.9.8.1735}
\showDOI{\tempurl}


\bibitem[\protect\citeauthoryear{Holmes and Spence}{Holmes and Spence}{2005}]%
        {MultisensoryIntegration}
\bibfield{author}{\bibinfo{person}{N.~P. Holmes} {and} \bibinfo{person}{C.
  Spence}.} \bibinfo{year}{2005}\natexlab{}.
\newblock \showarticletitle{Multisensory integration: Space, time and
  superadditivity}.
\newblock \bibinfo{journal}{\emph{” Current Biology}}  \bibinfo{volume}{15}
  (\bibinfo{date}{Spt} \bibinfo{year}{2005}), \bibinfo{pages}{R762–R764}.
\newblock


\bibitem[\protect\citeauthoryear{Kawano and Yanai}{Kawano and Yanai}{2014}]%
        {kawano14c}
\bibfield{author}{\bibinfo{person}{Y. Kawano} {and} \bibinfo{person}{K.
  Yanai}.} \bibinfo{year}{2014}\natexlab{}.
\newblock \showarticletitle{Automatic Expansion of a Food Image Dataset
  Leveraging Existing Categories with Domain Adaptation}. In
  \bibinfo{booktitle}{\emph{Proc. of ECCV Workshop on Transferring and Adapting
  Source Knowledge in Computer Vision (TASK-CV)}}.
\newblock


\bibitem[\protect\citeauthoryear{Kingma and Ba}{Kingma and Ba}{2014}]%
        {Adam}
\bibfield{author}{\bibinfo{person}{Diederik~P. Kingma} {and}
  \bibinfo{person}{Jimmy Ba}.} \bibinfo{year}{2014}\natexlab{}.
\newblock \bibinfo{title}{Adam: A Method for Stochastic Optimization}.
\newblock
\newblock
\urldef\tempurl%
\url{http://arxiv.org/abs/1412.6980}
\showURL{%
\tempurl}
\newblock
\shownote{cite arxiv:1412.6980Comment: Published as a conference paper at the
  3rd International Conference for Learning Representations, San Diego, 2015.}


\bibitem[\protect\citeauthoryear{Kiros, Zhu, Salakhutdinov, Zemel, Torralba,
  Urtasun, and Fidler}{Kiros et~al\mbox{.}}{2015}]%
        {DBLP:journals/corr/KirosZSZTUF15}
\bibfield{author}{\bibinfo{person}{Ryan Kiros}, \bibinfo{person}{Yukun Zhu},
  \bibinfo{person}{Ruslan Salakhutdinov}, \bibinfo{person}{Richard~S. Zemel},
  \bibinfo{person}{Antonio Torralba}, \bibinfo{person}{Raquel Urtasun}, {and}
  \bibinfo{person}{Sanja Fidler}.} \bibinfo{year}{2015}\natexlab{}.
\newblock \showarticletitle{Skip-Thought Vectors}.
\newblock \bibinfo{journal}{\emph{CoRR}}  \bibinfo{volume}{abs/1506.06726}
  (\bibinfo{year}{2015}).
\newblock
\showeprint[arxiv]{1506.06726}
\urldef\tempurl%
\url{http://arxiv.org/abs/1506.06726}
\showURL{%
\tempurl}


\bibitem[\protect\citeauthoryear{Lu, Allegra, Anthimopoulos, Stanco, Farinella,
  and Mougiakakou}{Lu et~al\mbox{.}}{2018}]%
        {DBLP:journals/corr/abs-1806-10343}
\bibfield{author}{\bibinfo{person}{Ya Lu}, \bibinfo{person}{Dario Allegra},
  \bibinfo{person}{Marios Anthimopoulos}, \bibinfo{person}{Filippo Stanco},
  \bibinfo{person}{Giovanni~Maria Farinella}, {and}
  \bibinfo{person}{Stavroula~G. Mougiakakou}.} \bibinfo{year}{2018}\natexlab{}.
\newblock \showarticletitle{A Multi-Task Learning Approach for Meal
  Assessment}.
\newblock \bibinfo{journal}{\emph{CoRR}}  \bibinfo{volume}{abs/1806.10343}
  (\bibinfo{year}{2018}).
\newblock
\showeprint[arxiv]{1806.10343}
\urldef\tempurl%
\url{http://arxiv.org/abs/1806.10343}
\showURL{%
\tempurl}


\bibitem[\protect\citeauthoryear{Mar{\'{\i}}n, Biswas, Ofli, Hynes, Salvador,
  Aytar, Weber, and Torralba}{Mar{\'{\i}}n et~al\mbox{.}}{2018}]%
        {Recipe1M}
\bibfield{author}{\bibinfo{person}{Javier Mar{\'{\i}}n},
  \bibinfo{person}{Aritro Biswas}, \bibinfo{person}{Ferda Ofli},
  \bibinfo{person}{Nicholas Hynes}, \bibinfo{person}{Amaia Salvador},
  \bibinfo{person}{Yusuf Aytar}, \bibinfo{person}{Ingmar Weber}, {and}
  \bibinfo{person}{Antonio Torralba}.} \bibinfo{year}{2018}\natexlab{}.
\newblock \showarticletitle{Recipe1M: {A} Dataset for Learning Cross-Modal
  Embeddings for Cooking Recipes and Food Images}.
\newblock \bibinfo{journal}{\emph{CoRR}}  \bibinfo{volume}{abs/1810.06553}
  (\bibinfo{year}{2018}).
\newblock
\showeprint[arxiv]{1810.06553}
\urldef\tempurl%
\url{http://arxiv.org/abs/1810.06553}
\showURL{%
\tempurl}


\bibitem[\protect\citeauthoryear{Matsuda, Hoashi, and Yanai}{Matsuda
  et~al\mbox{.}}{2012}]%
        {matsuda12}
\bibfield{author}{\bibinfo{person}{Y. Matsuda}, \bibinfo{person}{H. Hoashi},
  {and} \bibinfo{person}{K. Yanai}.} \bibinfo{year}{2012}\natexlab{}.
\newblock \showarticletitle{Recognition of Multiple-Food Images by Detecting
  Candidate Regions}. In \bibinfo{booktitle}{\emph{Proc. of IEEE International
  Conference on Multimedia and Expo (ICME)}}.
\newblock


\bibitem[\protect\citeauthoryear{Mikolov, Sutskever, Chen, Corrado, and
  Dean}{Mikolov et~al\mbox{.}}{2013}]%
        {Mikolov:2013:DRW:2999792.2999959}
\bibfield{author}{\bibinfo{person}{Tomas Mikolov}, \bibinfo{person}{Ilya
  Sutskever}, \bibinfo{person}{Kai Chen}, \bibinfo{person}{Greg Corrado}, {and}
  \bibinfo{person}{Jeffrey Dean}.} \bibinfo{year}{2013}\natexlab{}.
\newblock \showarticletitle{Distributed Representations of Words and Phrases
  and Their Compositionality}. In \bibinfo{booktitle}{\emph{Proceedings of the
  26th International Conference on Neural Information Processing Systems -
  Volume 2}} \emph{(\bibinfo{series}{NIPS'13})}. \bibinfo{publisher}{Curran
  Associates Inc.}, \bibinfo{address}{USA}, \bibinfo{pages}{3111--3119}.
\newblock
\urldef\tempurl%
\url{http://dl.acm.org/citation.cfm?id=2999792.2999959}
\showURL{%
\tempurl}


\bibitem[\protect\citeauthoryear{of~Agriculture}{of~Agriculture}{2019}]%
        {FoodData}
\bibfield{author}{\bibinfo{person}{U.S.~Department of Agriculture}.}
  \bibinfo{year}{2019}\natexlab{}.
\newblock \bibinfo{title}{Agricultural Research Service. FoodData Central}.
\newblock
\newblock


\bibitem[\protect\citeauthoryear{{Pan}, {Pouyanfar}, {Chen}, {Qin}, and
  {Chen}}{{Pan} et~al\mbox{.}}{2017}]%
        {8181494}
\bibfield{author}{\bibinfo{person}{L. {Pan}}, \bibinfo{person}{S. {Pouyanfar}},
  \bibinfo{person}{H. {Chen}}, \bibinfo{person}{J. {Qin}}, {and}
  \bibinfo{person}{S. {Chen}}.} \bibinfo{year}{2017}\natexlab{}.
\newblock \showarticletitle{DeepFood: Automatic Multi-Class Classification of
  Food Ingredients Using Deep Learning}. In \bibinfo{booktitle}{\emph{2017 IEEE
  3rd International Conference on Collaboration and Internet Computing (CIC)}}.
  \bibinfo{pages}{181--189}.
\newblock
\urldef\tempurl%
\url{https://doi.org/10.1109/CIC.2017.00033}
\showDOI{\tempurl}


\bibitem[\protect\citeauthoryear{Salvador, Hynes, Aytar, Marin, Ofli, Weber,
  and Torralba}{Salvador et~al\mbox{.}}{2017}]%
        {Salvador}
\bibfield{author}{\bibinfo{person}{A. Salvador}, \bibinfo{person}{N. Hynes},
  \bibinfo{person}{Y. Aytar}, \bibinfo{person}{J. Marin}, \bibinfo{person}{F.
  Ofli}, \bibinfo{person}{I. Weber}, {and} \bibinfo{person}{A. Torralba}.}
  \bibinfo{year}{2017}\natexlab{}.
\newblock \showarticletitle{Learning Cross-Modal Embeddings for Cooking Recipes
  and Food Images}. In \bibinfo{booktitle}{\emph{2017 IEEE Conference on
  Computer Vision and Pattern Recognition (CVPR)}}.
  \bibinfo{pages}{3068--3076}.
\newblock
\showISSN{1063-6919}
\urldef\tempurl%
\url{https://doi.org/10.1109/CVPR.2017.327}
\showDOI{\tempurl}


\bibitem[\protect\citeauthoryear{van~der Maaten and Hinton}{van~der Maaten and
  Hinton}{2008}]%
        {Maaten08visualizingdata}
\bibfield{author}{\bibinfo{person}{Laurens van~der Maaten} {and}
  \bibinfo{person}{Geoffrey Hinton}.} \bibinfo{year}{2008}\natexlab{}.
\newblock \bibinfo{title}{Visualizing data using t-SNE}.
\newblock
\newblock


\bibitem[\protect\citeauthoryear{Vaswani, Shazeer, Parmar, Uszkoreit, Jones,
  Gomez, Kaiser, and Polosukhin}{Vaswani et~al\mbox{.}}{2017}]%
        {DBLP:journals/corr/VaswaniSPUJGKP17}
\bibfield{author}{\bibinfo{person}{Ashish Vaswani}, \bibinfo{person}{Noam
  Shazeer}, \bibinfo{person}{Niki Parmar}, \bibinfo{person}{Jakob Uszkoreit},
  \bibinfo{person}{Llion Jones}, \bibinfo{person}{Aidan~N. Gomez},
  \bibinfo{person}{Lukasz Kaiser}, {and} \bibinfo{person}{Illia Polosukhin}.}
  \bibinfo{year}{2017}\natexlab{}.
\newblock \showarticletitle{Attention Is All You Need}.
\newblock \bibinfo{journal}{\emph{CoRR}}  \bibinfo{volume}{abs/1706.03762}
  (\bibinfo{year}{2017}).
\newblock
\showeprint[arxiv]{1706.03762}
\urldef\tempurl%
\url{http://arxiv.org/abs/1706.03762}
\showURL{%
\tempurl}


\bibitem[\protect\citeauthoryear{Wang, Kumar, Thome, Cord, and Precioso}{Wang
  et~al\mbox{.}}{2015}]%
        {Wang}
\bibfield{author}{\bibinfo{person}{Xin Wang}, \bibinfo{person}{D. Kumar},
  \bibinfo{person}{N. Thome}, \bibinfo{person}{M. Cord}, {and}
  \bibinfo{person}{F. Precioso}.} \bibinfo{year}{2015}\natexlab{}.
\newblock \showarticletitle{Recipe recognition with large multimodal food
  dataset}. In \bibinfo{booktitle}{\emph{2015 IEEE International Conference on
  Multimedia Expo Workshops (ICMEW)}}. \bibinfo{pages}{1--6}.
\newblock
\urldef\tempurl%
\url{https://doi.org/10.1109/ICMEW.2015.7169757}
\showDOI{\tempurl}


\bibitem[\protect\citeauthoryear{Xu, Herranz, Jiang, Wang, Song, and Jain}{Xu
  et~al\mbox{.}}{2015}]%
        {GeolocalizedModelingforDishRecognition}
\bibfield{author}{\bibinfo{person}{R. Xu}, \bibinfo{person}{L. Herranz},
  \bibinfo{person}{S. Jiang}, \bibinfo{person}{S. Wang}, \bibinfo{person}{X.
  Song}, {and} \bibinfo{person}{R. Jain}.} \bibinfo{year}{2015}\natexlab{}.
\newblock \showarticletitle{Geolocalized Modeling for Dish Recognition}.
\newblock \bibinfo{journal}{\emph{IEEE Transactions on Multimedia}}
  \bibinfo{volume}{17}, \bibinfo{number}{8} (\bibinfo{date}{Aug}
  \bibinfo{year}{2015}), \bibinfo{pages}{1187--1199}.
\newblock
\showISSN{1520-9210}
\urldef\tempurl%
\url{https://doi.org/10.1109/TMM.2015.2438717}
\showDOI{\tempurl}


\bibitem[\protect\citeauthoryear{{Yanai} and {Kawano}}{{Yanai} and
  {Kawano}}{2015}]%
        {7169816}
\bibfield{author}{\bibinfo{person}{K. {Yanai}} {and} \bibinfo{person}{Y.
  {Kawano}}.} \bibinfo{year}{2015}\natexlab{}.
\newblock \showarticletitle{Food image recognition using deep convolutional
  network with pre-training and fine-tuning}. In \bibinfo{booktitle}{\emph{2015
  IEEE International Conference on Multimedia Expo Workshops (ICMEW)}}.
  \bibinfo{pages}{1--6}.
\newblock
\urldef\tempurl%
\url{https://doi.org/10.1109/ICMEW.2015.7169816}
\showDOI{\tempurl}


\end{thebibliography}

\end{document}